\documentclass[conference,10pt]{IEEEtran}
\IEEEoverridecommandlockouts
\usepackage{cite}
\usepackage{amsmath,amssymb,amsfonts}
\usepackage{algorithmic}
\usepackage{graphicx}
\usepackage{textcomp}
\usepackage{xcolor}
\def\BibTeX{{\rm B\kern-.05em{\sc i\kern-.025em b}\kern-.08em
    T\kern-.1667em\lower.7ex\hbox{E}\kern-.125emX}}

\usepackage{changes}
\definechangesauthor[color=orange]{rs}
\definechangesauthor[color=red]{pr}

\begin{document}

% \title{GitHub Mining of Quantum Computing Contributors}
\title{Making Quantum Computing Open: Lessons from Open-Source Projects}

\author{
\IEEEauthorblockN{Ruslan Shaydulin, Caleb Thomas, and Paige Rodeghero\\}
\IEEEauthorblockA{School of Computing\\
Clemson University\\
Clemson, SC, USA\\
Email: \{rshaydu, caleb8, prodegh\}@g.clemson.edu}
}

\maketitle

\begin{abstract}
Quantum computing (QC) is an emerging computing paradigm with potential to revolutionize the field of computing. QC is a field that is quickly developing globally and has high barriers of entry. In this paper we explore both successful contributors to the field as well as wider QC community with the goal of understanding the backgrounds and training that helped them succeed. We gather data on 148 contributors to open-source quantum computing projects hosted on GitHub and survey 46 members of QC community. Our findings show that QC practitioners and enthusiasts have diverse backgrounds, with most of them having a PhD and trained in physics or computer science. We observe a lack of educational resources on quantum computing. Our goal for these findings is to start a conversation about how best to prepare the next generation of QC researchers and practitioners.
\end{abstract}

\section{Introduction}

The recent years saw a sequence of rapid scientific and engineering advancements in the field of quantum computing (QC). In less than two decades QC evolved from being mostly theoretical to being able to solve practically interesting problems~\cite{shaydulin2018network, shaydulin2018community,farhi2014quantumbounded, ushijima2017graph}. A number of private companies, universities and government labs worldwide successfully demonstrated a hardware implementation of QC~\cite{ballance2016high,barends2014superconducting, petta2005coherent, acin2018quantum,saffman2010quantum}, and a number of~quantum computing projects have emerged, aiming to provide a platform for integration of quantum computation in scientific and business workflows.

Quantum computing has the potential to provide exponential speedups to many important problems~\cite{nielsen2002quantum}, % including 
making possible to compute things that are unimaginable today. The most famous example of the potential capabilities of quantum computing is Shor's~\cite{shor1994algorithms} algorithm. It describes a way to factor integers in polynomial time 
%Shor's algorithm for  integer factoring~\cite{shor1994algorithms} with 
and has implications for RSA-based cybersecurity~\cite{nielsen2002quantum}.
Many more algorithms have been developed, with applications to quantum chemistry~\cite{romero2018strategies}, combinatorial optimization~\cite{farhi2014quantum,farhi2014quantumbounded} and machine learning~\cite{shaydulin2018network}. 

The key problem in the field of QC software engineering is that to fully realize the promise of quantum computing the community needs scientists and developers with skills relevant to the evolving field. In these early days, researchers working in quantum computing need an understanding of both the unique logic of quantum computation, as well as the computer science experience needed to integrate it into existing classical workflows. One example of an area where a diverse set of skills spanning multiple fields is required is development of quantum algorithms. Quantum algorithms like Shor's~\cite{shor1994algorithms} are significantly different from their classical counterparts. Just understanding them requires knowledge of quantum mechanics and computational complexity theory, two disciplines that usually do not go together in a university curriculum. Development of new algorithms requires mastery of both, as well as a deep computer science background. 

Our data shows that currently most QC practitioners deal with the problem of lack of educational resources by combining their formal training in fields like physics or computer science with reading papers and textbooks on their own and following tutorials online. This makes the field of QC hard to enter and hard to navigate for the newcomers. We address this problem by performing an analysis of the background and skills that practitioners find important to their work. To our knowledge this is the first attempt to explore quantum computing from the software engineering perspective.

In this paper, we present the data about the background and training of QC researchers and practitioners. The methodology that we follow consists of two parts, with possible overlap between the two. First, we scrape the data about 148 contributors to open-source QC projects from GitHub. Second, we survey 46 QC professionals to augment the data. We found that most QC specialists are trained in non-computing fields like physics, chemistry or mathematics. Our goals for this paper is to receive community feedback on this line of research which could include further investigation of factors contributing to success in QC field through surveying of practitioners, panels at conferences and gathering data on open-source contributions. We believe our findings can help develop educational resources targeted at preparing a new generation of~QC researchers and practitioners. %\looseness=-2

To the best of our knowledge, this is the first exploration of the quantum computing field from the software engineering perspective, and the first attempt to rigorously explore the contributors to open-source quantum computing projects. This makes our approach novel and interesting to both software engineering and quantum computing communities.

\section{Problem}

Today there are very few training programs in quantum computing and there is no consensus on the curriculum structure. This is supported by our data: out of 46 survey respondents, only one received formal training in quantum computation. This indicates a lack of understanding of what training is required for success in QC field. Without looking further into what is required to become a part of the QC field, it is difficult for aspiring QC researchers to know what they should do to become a part of this field, which will inevitably stunt the growth of the QC industry.  There is a need for a deeper exploration of the challenges facing QC researchers and practitioners.  This paper addresses this by exploring the contributors to open-source quantum computing projects, their backgrounds, and the challenges they face.

\section{Background and Related Work}

Quantum computing is still a very young field, with software projects constantly pushing the boundary of human knowledge. As such, quantum computing software projects have to deal with all the problems that plague scientific software projects: unforeseen changes in the requirements, lack of software development expertise and limited budgets~\cite{letondal2003anticipating}. The cross-disciplinary nature of quantum computing adds to the complexity of the domain.
% \\\\
% \vspace{1.0em}
\subsection{Quantum Computing}
% \noindent\textit{\textbf} 
 
Unlike in classical computation, where the computation happens by manipulating bits, the fundamental computational unit in QC is qubit. A bit can have one of two states: 0 or 1. Similarly, qubit state is a unit vector in a two-dimensional complex vector space~\cite{nielsen2002quantum}. A qubit state can be encoded in a state of a quantum mechanical object, for example as polarization of a single photon~\cite{nielsen2002quantum}.
 
The field of quantum computing is in a state of constant change, and is generally expected to continue changing in the foreseeable future. In the past few years multiple Near-term Intermediate-Scale Quantum (NISQ) hardware implementations have been developed~\cite{preskill2018quantum} and demonstrated to provide a potential for quantum speedups~\cite{dunjko2018computational}. Naturally, different implementations come with certain trade-offs. For example, trapped ion qubits are generally less noisy and offer better connectivity, whereas superconducting qubits offer faster gate clock speeds and more clear path to scalability~\cite{linke2017experimental}. This diversity of hardware introduces an additional degree of complexity for the development of QC algorithms and software, forcing algorithm developers to stay aware of the trade-offs presented by hardware.

A plethora of algorithms leveraging the power of quantum computation have been developed over the years. Shor's~\cite{shor1994algorithms} and Grover's~\cite{grover1996fast} algorithms are two most well-known examples of quantum algorithms for practical problems with theoretically proven speed-ups over classical state-of-the-art. However, the limitations of NISQ-era hardware make most of them impossible to run in the near-term. Near-term quantum computers are widely believed to be able to provide no more than a few hundreds of non error-corrected qubits. To address this challenge, a number of NISQ approaches have been proposed, most prominent of them Variational Quantum Eigensolver (VQE)~\cite{peruzzo2014variational} and Quantum Approximate Optimization Algorithm (QAOA)~\cite{farhi2014quantum}. The limitations of near-term hardware make development of practical algorithms especially challenging.

% \noindent\textit{\textbf{Open-source Scientific Software Projects}} 
\subsection{Open-source Scientific Software Projects}
% 	\vspace{1.0em}\\
    There has been a push towards open source scientific software projects.  Government labs, such as Sandia National Laboratories, create and use open source projects~\cite{WinNT, WinNT2}.  This allows for easier collaboration with  scientists and programmers both within industry and academia.   
    
    Many software engineering studies have been conducting by mining GitHub~\cite{kalliamvakou2016depth, cosentino2016findings, kalliamvakou2014promises, pletea2014security,vasilescu2015gender}.  These projects look at contributors, their efforts, the number of bugs created or fixed, etc. One problem with open source projects is that it can be unclear who is contributing more than others~\cite{Goldman:2011:RCC:2047196.2047215}.  Today, conferences such as ``The Mining Software Repositories'' (MSR) exist to better understand software repositories and the contributors.    
    
    %Quantum computing open source repositories are starting to appear on GitHub. Scientists interested in developing these repositories are at a disadvantage.  There are only a few people   
    
   % A large quantity of software projects are done in an open-source manner through services such as Github.  This recent trend makes an investigation of networkings on social coding sites valuable~\cite{letondal2003anticipating,Goldman:2011:RCC:2047196.2047215}.  However it is noteworthy that this process undermines the individual programmer's contribution to a product and it can sometimes be unclear who is contributing more than others~\cite{Goldman:2011:RCC:2047196.2047215}.  That is the reason why, along with mining for data contribution in Github, surveys are also be sent to the contributors in Github to get their feedback. More importantly, it is pertinent that the development of software that could support quantum computing devices is developed at least apace that of the development of said devices~\cite{ladd2010quantum, linke2017experimental}.  Otherwise, without those who would be writing the software, the hardware will be wasted without the proper handling via appropriate software~\cite{linke2017experimental}.  

\section{Methodology}

In order to understand the structure of quantum computing community and the challenges facing the contributors, we  focus on three open-source and open-development quantum computing projects from three companies working in quantum computing field: Qiskit (IBM), PyQuil/Grove (Rigetti) and Cirq (Google, not an official product).

Qiskit is open-source project developed by IBM. At the time of writing, it consists of two main parts: Qiskit Terra~\cite{qiskit-terra} provides the basic building blocks and Qiskit Aqua~\cite{qiskit-aqua} provides a library of algorithms upon which applications can be built. Qiskit is in active development and new modules with new functionality has been integrated into it while this paper was in preparation.

PyQuil/Grove is an open-source QC framework developed by Rigetti. Rigetti has recently announced Quantum Cloud Services (QCS), a new closed-source project aimed at providing cloud access to quantum computers to researchers and developers. Rigetti QCS is currently in closed beta-testing. In this work we focus on PyQuil/Grove, previous open-source iteration of Rigetti quantum effort. Rigetti PyQuil~\cite{pyquil} provides the low-level building blocks and Rigetti Grove~\cite{grove} provide a library of algorithms implemented with PyQuil. 

Cirq~\cite{cirq} is developed by researchers at Google. Cirq is not an official Google product. Cirq provides low-level tools for building programs to be run on noisy intermediate-scale quantum (NISQ) computers.

% Qiskit is open-source project developed by IBM. It consists of two main parts: Qiskit Terra~\cite{qiskit-terra} provides the basic building blocks and Qiskit Aqua~\cite{qiskit-aqua} provides a library of algorithms upon which applications can be built. PyQuil/Grove is an open-source QC framework developed by Rigetti. Rigetti PyQuil~\cite{pyquil} provides the low-level building blocks and Rigetti Grove~\cite{grove} provides a library of algorithms implemented with PyQuil. Cirq~\cite{cirq} is developed by researchers at Google. Cirq is not an official Google product. Cirq provides low-level tools for building programs to be run on noisy intermediate-scale quantum (NISQ) computers.

Our approach consists of two parts. First, we scraped the contribution data from the projects' GitHub repositories using PyDriller framework~\cite{PyDriller}. These data  provide us with a quantitative understanding of who contributes to the quantum computing projects. QC field is still relatively small when compared with software development as a whole. Analyzing the data on between a hundred and two hundred QC project contributors provides a valuable insight about the QC practitioners. Second, we augment the data from by sending out a questionnaire to quantum computing developers and enthusiasts through IBM Qiskit and Rigetti Slack channels, as well as to emails of contributors we scraped from GitHub. 

The questionnaire  extends the data harvested from GitHub by exploring the background, motivation and the challenges facing the practitioners. %Questions include simple ones like "What was your major in college?", as well as more open-ended ones like "What CS class you wish you took but didn't?". 
The questionnaire consists of three parts. First part includes questions on the participants background, education and professional experience (e.g. "What was your major in college?"). Second part explores participants'  experience with different QC frameworks, what educational resources they find the most helpful and how confident they are in their CS  and physics knowledge (e.g. "How adequate do you find your computer science (CS) background for your work in QC (e.g. coding skills,
understanding of CS concepts etc)?"). Third part consists of open-ended questions, e.g. "What kind of CS training would have better prepared you for your work in QC? What CS class you wish you took but didn’t?".
%as well as more open-ended questions like "What kind of CS training would have better prepared you for your work in QC? What CS class you wish you took but didn’t?".

%What kind of CS training would have better prepared you for your work in QC?

% \begin{itemize}
% \item What was your major in college?
% \item Do you have a graduate degree? If so, is it Master's or PhD and in what field?
% \end{itemize}

% \noindent as well as more open-ended ones like:

% \begin{itemize}
% \item How did you become involved in quantum computing?
% \item What part of the work that you do in QC research / development, do you find the most \mbox{challenging}?
% \end{itemize}

\section{Results}

We collected data about 148 quantum computing researchers by scraping data on contributions from GitHub repositories of popular QC projects and received 46 responses to our survey. We can say that the first part of the data (GitHub data) give us an insight about the people who \emph{already contribute} to quantum computing projects, whereas the second part of the data (survey data) gives us a broader picture of the community of people \emph{interested} in quantum computing. We observe a significant difference between the two, indicating that QC companies need to engage the broader community in software development process to make these projects truly collaborative. The GitHub data also demonstrates how different companies approach the challenging task of building a quantum computing framework in different ways.

Through analysis of the GitHub data we get an idea of people who are already successful in the field. Concretely, we can measure the success by number of commits to the state-of-the-art open source projects. If success is defined like that, the data shows that most successful contributors have a PhD. 61\% of contributors with more than 10 commits in repository's master branch have a PhD, with 81\% of them having a PhD in physics (55\%) or computer science (26\%). This is unsurprising, since the projects in questions are fundamentally research projects, requiring precisely the skills a PhD program develops. GitHub data demonstrates how the company culture varies between the three big players: while in IBM and Rigetti projects 62\% and 64\% top contributors have a PhD, for Google's Cirq the number is only 50\%, indicating that Google treats its quantum computing initiative differently. What unites all the projects we looked at is that they were all developed almost exclusively by people employed at that company, with the ratio of employees among top contributors being 100\% for all three companies. This indicates a need for a broader collaboration, both between companies, as well as between companies and other research institutions.  

In contrast to the GitHub data, which contains almost exclusively employees of the QC companies, out of 64 survey respondents only 30\% indicated that they work for a quantum-computing-centered company. This shows that there is a broader interest in quantum computing, both in industry (35\% of respondents) as well as in national laboratories (9\%) and among PhD students (9\%). Still, we observe a very high number of people with PhD among our respondents (28\%), confirming our observations from GitHub data. Similarly to GitHub data, most respondents (70\%) studied either physics (29\%) or computer science (41\%). 

The predominance of people with PhDs (i.e. people with research backgrounds) is indicative of a young field, where the barrier to entry is still very high. There is a clear need to lower those barriers by developing educational materials that could help bring people into the field. The lack of educational materials is confirmed by multiple observations. First, our data shows almost no people with degree in Quantum Computation, Quantum Information or related fields (only one respondent indicated that they have  a masters degree in Quantum Computing). Second, in ranking educational materials by their impact and helpfulness the respondents rate "reading papers on their own" (mean score of 2.98, less is better) and "reading textbooks on their own" (mean score of 3.04) as most helpful, followed by "following tutorials in software repositories" (mean score of 3.25). This shows that the effort software companies have put into developing introductory tutorials to accompany their frameworks is paying off and that the community finds them helpful. Surprisingly, Massive Open Online Courses (MOOCs) scored fairly low (second to last with mean score of 3.91), which in our opinion says more about their accessibility rather than availability (that is how easy their are to follow, rather then how easy their are to find). The only online MOOCs known to us are parts of MIT Quantum Computing Curriculum, require a strong mathematical background and teach mostly theoretical (and somewhat challenging) material. 

The data we collected shows that training for quantum computing should combine physics, computer science (CS) and mathematics. We find that respondents trained in predominantly non-CS fields find their computer science skills lacking (skills like coding, software development etc), where\-as respondents with predominantly CS training find their physics knowledge inadequate. Interestingly, the ratio of respondents who found their physics skills moderately or extremely inadequate (20\% of respondents) is much higher that the ratio of respondents who said the same about their CS skills (9\%). This indicates that the community finds the physics side harder, suggesting that the training should focus more on developing relevant physics knowledge. In an open-ended question, where the respondents were encouraged to discuss what training they wish they received, 17 respondents indicated that they didn't receive sufficient training in quantum computing and quantum information. Another commonly received suggestion was linear algebra and mathematical training in general.  

% \vspace{-0.4cm}

\section{Discussion and Conclusion}

In this paper we present the data on 148 contributors to quantum computing projects and 46 members of wider quantum computing community (with possible overlap between the two). Two parts of our dataset complement each other, providing a valuable insight into that factors contributing to success in the field of quantum computing.

Our finding that most respondents rate studying on their own as the most useful highlights the need for universities and other educational institutions to step in and offer training, be it in form of a MOOC available on the internet or an in-person degree program. These educational resources should adequately prepare QC practitioners, with more focus given to the most challenging parts of QC curriculum.
Our findings have implications for designs of such offerings, indicating that the respondents find computer science and software engineering aspect of their work in QC the least challenging, while struggling more with underlying fundamental mathematical and physical concepts. 

The majority of the respondents have less than two years of experience in QC (80\%), with only 7\% having worked in the field for more than 5 years. This is indicative of a young and booming field, with a lot of new entrants. The relative immaturity of the field as well as its interdisciplinarity make it very challenging to join. As a community, we need to develop more resources to bringing quantum computing to undergraduate students through classes, online courses and tutorials. But as the field is booming and hype is high, it's important to keep a cool head. As one of our respondents wisely noted, one should "be careful not to train people for something that will bust and not get them jobs." 

% There are various articles depicting the impact that quantum computing would have on the computer field once it is developed.  In \emph{Quantum Computers}~\cite{ladd2010quantum}, the possible uses of quantum computing are discussed and it is concluded that the field of computer science field would be improved by further research and development in the quantum computing field.  Since it is still largely unknown who is currently working in the quantum computing industry, for this to become a reality it is adamant that the contributors to the field be named to speed up the process of research and development.  

We hope this project will have a positive impact by bringing software engineering community and the software engineering methods to quantum computing. Moreover, we believe that the data we collected will contribute to the discussion on how best to prepare the next generation of quantum computing specialists and researchers. 

\medskip

\noindent\textit{Note Added. --} After the completion of this work, we became aware of a related work appearing in Ref.~\cite{fingerhuth2018open}. They provide an exhaustive review of open-source quantum computing projects. Unlike our work, they don't directly survey the contributors to these projects. Additionally, we review the educational and professional background of the contributors.

\bibliographystyle{IEEEtran}
\bibliography{quantum-contributors,qaoa}

\end{document}